\documentclass[prl,preprint]{revtex4}
\usepackage{graphicx}
\usepackage{dcolumn}
\usepackage{bm}
\begin{document}
\title{Inertial Screening in Sedimentation}
\author{P.N. Segr{\` e}}
\address{Department of  Physics, Emory
University, Atlanta, GA 30322}
\begin{abstract}
We use particle image velocimetry to measure the sedimentation
dynamics of a semi-dilute suspension of non-Brownian  spheres at
Reynolds numbers, $0.001\le Re\le 2.3$, extending from the Stokes to
the moderately inertial regime. We find that the onset of inertial
corrections to Stokes sedimentation occurs when the inertial
screening length $l=a/Re$ becomes similar to the Stokes
sedimentation length $\xi_0$, at $Re_c= a/\xi_0\approx 0.05$. For
$Re>Re_c$, inertial screening significantly reduces both the
magnitude and spatial extent of the particle velocity  fluctuations.
  A modified Hinch force balance model
    connects the
    fluctuation magnitudes $\sigma_V/V$ to the correlation sizes
    $\xi$.

\end{abstract}
\maketitle

The sedimentation dynamics  of
   non-Brownian spheres is a fundamental problem in
   physics \cite{hinch} and has been the subject of intense activity in
   recent years (for a review see \cite{review1}).
   Much of the focus has been
  on  particles slowly settling in  viscous
liquids, conditions that correspond to very low Reynolds numbers,
$Re\ll 1$, called the Stokes regime. The Reynolds number is the
ratio of inertial to viscous forces in fluids, and is defined as
$Re\equiv 2Va/\nu$ ($V$ is the particle velocity, $a$ the radius,
and $\nu$ the fluid kinematic viscosity). In the Stokes regime,
where inertial forces are insignificant, concentrations of spheres
display $Re$-independent large amplitude velocity fluctuations
$\sigma_V/V$  during settling \cite{nandg1,
nandg2,segreprl,nature,michel}. Significantly, experiments
\cite{segreprl,nature,michel,bruce}, simulations \cite{tony1}, and
theory \cite{theory2} have shown that the fluctuations display  a
characteristic spatial size $\xi$, despite the fact that the
hydrodynamic interactions emanating from a single {\it isolated}
sphere are of infinite range ($\propto 1/r$) \cite{batch}. The
Stokes screening length $\xi$  represents regions of
 concentration fluctuations, $\sigma_\phi$, that drive
velocity fluctuations $\sigma_V$ and particle diffusion $D\sim
\sigma_V\xi$.
 While an explanation of the origin of  screening in Stokes sedimentation
 remains controversial,
its existence and central importance for a description of
sedimentation is not.

The sedimentation dynamics at higher speed flows where inertial
forces become significant,
 $Re\sim 1$, have received much less attention despite its fundamental
importance and widespread relevance to numerous chemical
   industries \cite{review}.
    In contrast to the Stokes regime, for an
    {\it isolated}  sphere falling at moderate
$Re$, there is a distance  beyond which the $1/r$ Stokes like
hydrodynamics are screened \cite{batch}. The inertial screening
length, $l\sim a/Re$, becomes shorter as the Reynolds number is
increased. For concentrations of settling spheres, the Reynolds
number dependence of the  hydrodynamic interactions brought about by
inertia is predicted to result in  Reynolds number dependent
sedimentation dynamics as well \cite{koch}. Experiments designed to
investigate these changes, however,  are greatly lacking. We are
aware of only a single experiment, by Cowan et {\it al.} \cite{john}
using novel ultrasonic techniques, that has examined the fluctuation
dynamics of spheres at moderate $Re\lesssim 1$. Surprisingly, they
concluded that the velocity fluctuations were independent of
Reynolds number for $Re\lesssim 1$. In the absence of experiments
describing the sedimentation dynamics beyond the Stokes regime, our
understanding of the basic physics, and our ability to develop and
test model theories, remains very limited.

In this letter, we describe experiments that demonstrate how
 moderate amounts of fluid inertia can significantly affect the settling
dynamics of spheres. The onset of inertial corrections to Stokes
sedimentation occurs when
    the inertial screening length $l=a/Re$ becomes as small as  the Stokes
    sedimentation length  $\xi_0$, at $Re_c= a/\xi_0$.
    For   $Re>Re_c$,
inertial screening significantly reduces both the magnitude and
spatial extent of the particle velocity
    fluctuations.  A modified Hinch force balance model,
    with Reynolds number dependent drag coefficients,
    connects the
    fluctuation magnitudes $\sigma_V/V$ to the correlation sizes $\xi$ over the
    entire  range studied, $0.001\le Re\le 2.3$.

    The particles used in our experiments were monodisperse glass
    beads of radius $a=137\pm 9$ $\mu$m.
    They were dispersed in various mixtures of glycerol
    and water,
    enabling us to  examine the very low to moderate
    Reynolds number regimes, $0.001\le Re
\le 2.3$. In all cases the volume fraction is $\phi=0.06$.
    The sample cell was a rectangular glass tube of dimension
    $8\times 80\times 305$ mm, and the temperature was at the ambient
    value $T=23\pm 1 ^0$C.
    Particle velocities were measured using a particle
image velocimetry (PIV) \cite{adrian} apparatus consisting of some
specialized image processing software and hardware purchased from
Dantec Instruments. A large cross section of the cell was imaged
($3\times 3$ cm), so that several thousand particles could be
simultaneously studied. Initially, random dispersions were prepared
by vigorous shaking of the cells. The location of the imaging window
was far from both the sedimentation front and the sediment growth.
Each velocity fields is a map of $41\times42$ vectors, and each
velocity vector is the local average of two to four spheres.

\begin{figure}\includegraphics[width=140mm]{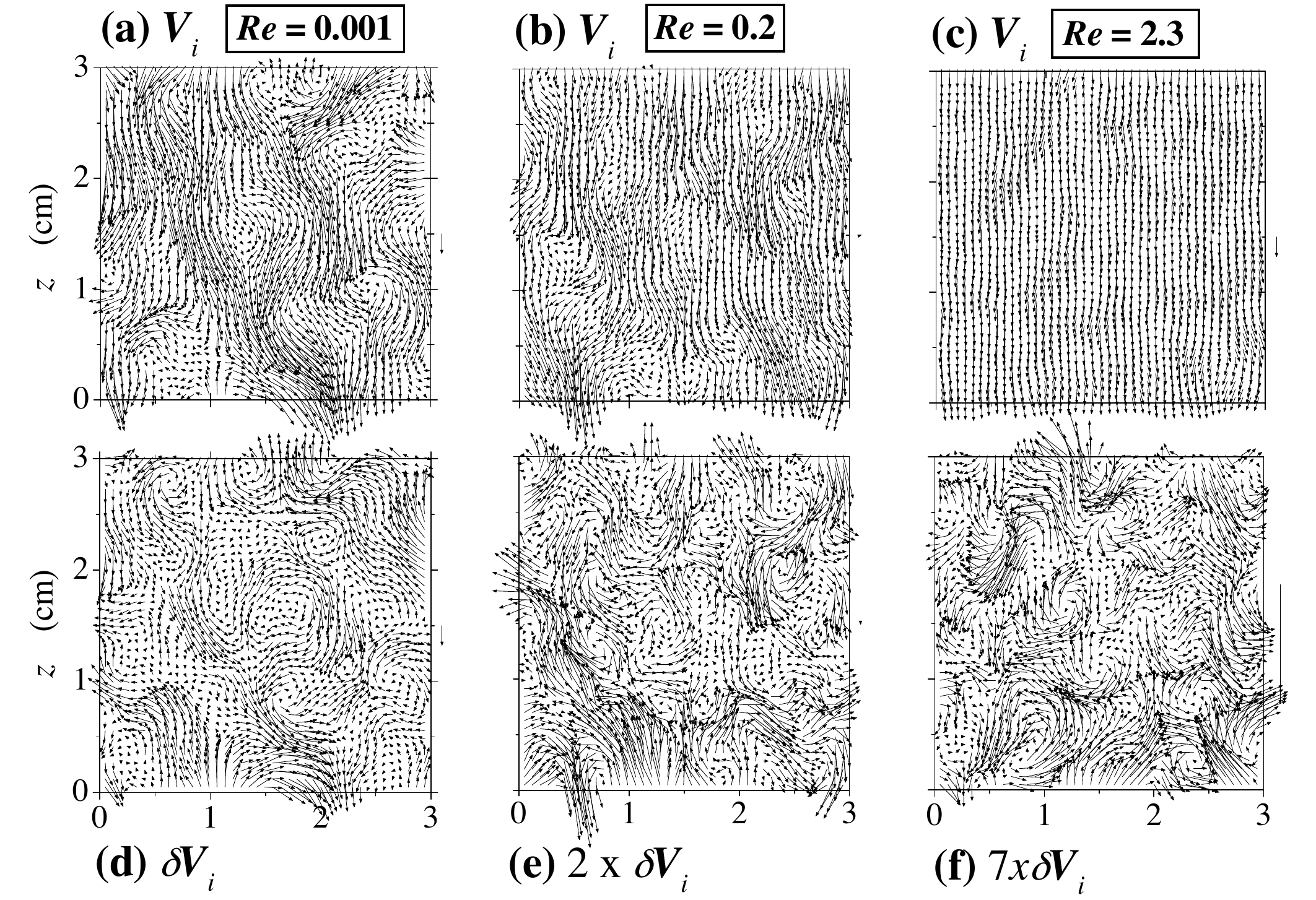}
 \caption{PIV results for three suspensions of
sedimenting spheres at $\phi=0.06$ and Reynolds numbers (a,d)
$Re=0.001$,  (b,e) $Re=0.2$, and  (c,f) $Re=2.3$. (a-c) Velocities
${\bf V}_i$.
 (d-f)  Velocity fluctuations
calculated from (a-c) using $\delta {\bf V_i}={\bf V_i}-\langle {\bf
 V_i} \rangle$. (Note the magnified scales in (e,f)). The single
vector to the right of each map represents  the mean velocity
$\langle {\bf
 V_i} \rangle$. } \label{fig:map}
\end{figure}

We begin by examining   typical particle velocity  fields during
sedimentation. Figures \ref{fig:map}(a-c) show results for particle
velocities ${\bf V}_i$ from three samples spanning the very low,
$Re=0.001\ll 1$, to moderate, $Re=2.3$, Reynolds number regimes.
In the Stokes flow regime,
     Fig. \ref{fig:map}(a), the patterns look similar to those
     reported in the literature \cite{segreprl,michel},
     with large magnitude fluctuations
     occurring in extended {\it swirling} regions.
 In the presence of some inertia,
$Re=0.2$     in Fig. \ref{fig:map}(b),   the pattern appears
slightly more uniform, indicating a slight reduction in the
fluctuations relative to that seen at $Re=0.001$. At the highest
flow speed,
     $Re=2.3$ in Fig. \ref{fig:map}(c), there is a dramatic change
     in the dynamics.  The fluctuations  are largely
absent,
      the particles appear to
     be nearly  falling with the same speed. We interpret these changes as  evidence of
     an inertial screening of the velocity fluctuations.

     To examine the fluctuations in more detail,
      Figs. \ref{fig:map}(d-f)  shows the same set of maps
     with the mean velocities subtracted off,  $\delta {\bf
V}_i\equiv{\bf V}_i-\langle {\bf V}_i\rangle$. All of the patterns
in fact show qualitatively similar swirling regions. The sizes of
the regions appear to decrease with $Re$. The dramatic decrease in
fluctuations seen in Fig. \ref{fig:map}(c) is reflected in the
greatly magnified scale ($7\times \delta {\bf V_i})$ of Fig.
\ref{fig:map}(f) relative to Fig. \ref{fig:map}(d).

 By collecting large numbers of velocity maps  over the majority of each falling column,
  we can access ensemble averaged information.
    We first examine the spatial correlations of the velocity fluctuations.
     The normalized autocorrelation function of
the $z$ component ($\|$ to gravity) of the velocity fluctuations is
defined as $C_z({\bf r})\equiv \langle \delta V_z(0)\delta V_z({\bf
r})\rangle/\langle \delta V_z(0)^2\rangle$, where $\langle \ldots
\rangle$ represents an ensemble average of several hundred vector
maps. The distance vector ${\bf r}$ is either in the direction
parallel to gravity, $C_z(z)$, or perpendicular to it, $C_z(x)$.

\begin{figure}
\includegraphics[width=84mm]{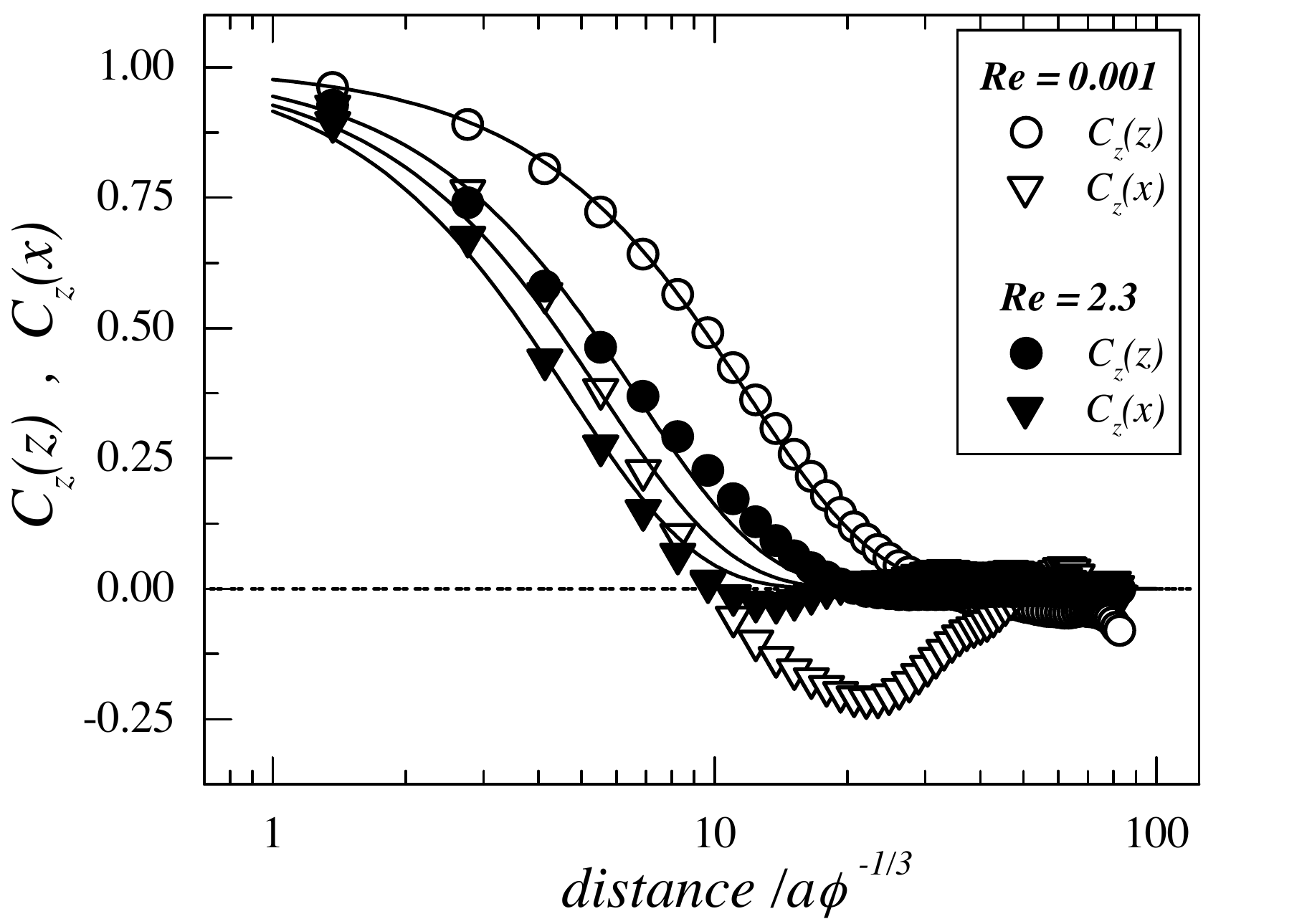}
\caption{Spatial correlation functions of the $z$ component of the
velocity fluctuations as a function of distance  $\|$, $C_z(z)$, and
$\perp$, $C_z(x)$, to the $z-$axis for $Re=0.001$ and $Re=2.3$. }
\label{fig:cf}
\end{figure}

To search for evidence of inertial screening in the spatial velocity
correlations, Fig. \ref{fig:cf} compares the correlation functions
between a sample in the Stokes regime, $Re=0.001$, and one with
significant inertia, $Re=2.3$. In all cases the functions decay to
zero at large distances, indicative of finite range correlations.
The decay lengths, however, are much reduced in the inertial sample.
The form of the perpendicular correlation function is also affected,
the long-range negative correlation dip in $C_z(x)$  is completely
absent in the inertial case.

\begin{figure}[t]\includegraphics[width=84mm]{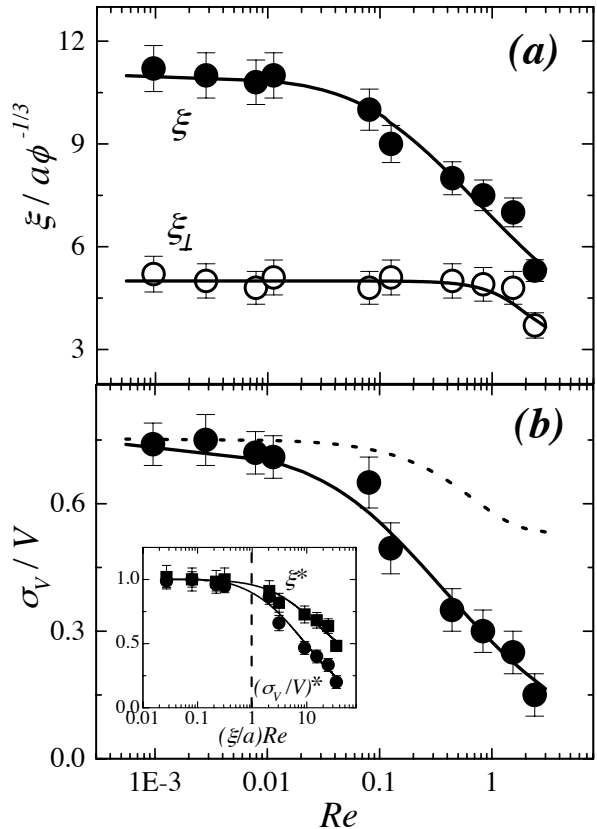}
\caption{Inertial screening of velocity fluctuations. (a) Spatial
correlation lengths $\xi$ and $\xi_\perp$ as a function of Reynolds
number $Re$. The line through $\xi$ is the empirical fit Eq.
(\ref{eq:xifit}). The line through $\xi_\perp$ is a guide to the
eye. (b) Normalized velocity fluctuations $\sigma_V/V$ vs.  $Re$.
The solid line is  the Hinch model Eq. (\ref{eq:dvre}). The dashed
line is Eq. (\ref{eq:dvre}) with $Re=0$. Inset: $\xi$ and
$(\sigma_V/V)$, normalized by their Stokes values, vs. $(\xi/a)Re$.}
\label{fig:xidv}
\end{figure}
To quantify the range of the velocity correlation functions, we fit
to the empirical forms $C_z(z)=\exp[-(z/\xi)^{1.5}]$ and
$C_z(x)=\exp[-(x/\xi_\perp)^{1.5}]$, as shown in Fig. \ref{fig:cf}.
Results are given in Fig. \ref{fig:xidv}(a) for $\xi$ and
$\xi_\perp$ from all of our samples ranging from $0.001\le Re\le
2.3$. The longitudinal length $\xi$, the longer of the two, shows
the most change with $Re$. For $Re\lesssim 0.05$, $\xi$ is
independent of $Re$ as expected in the Stokes regime, and the values
are all in good agreement with the scaling relation $\xi_0\approx
11a\phi^{-1/3}$ previously found in very low Reynolds number
sedimentation \cite{segreprl}. For $Re\gtrsim 0.05$, the behavior
begins to change, and $\xi$ follows a roughly logarithmic decay with
$Re$. The decrease of $\xi$ with $Re$ demonstrates that inertial
screening occurs at shorter  distances for higher Reynolds numbers.
In agreement with this, the short range transverse correlations
don't show any inertial influence until much higher Reynolds
numbers, $Re=2.3$. For analysis purposes below, we note that $\xi$
can be well fit by
\begin{equation}
\xi \simeq \xi_0/(1+5.5Re)^{1/4} \;. \label{eq:xifit}
\end{equation}

The velocity maps in Fig. \ref{fig:map} also show dramatic changes
to the {\it magnitudes} of the velocity fluctuations as $Re$ is
increased. To quantify this we consider the
  ensemble averaged {\it rms} velocity
fluctuations,  $\sigma_v \equiv \sqrt{\langle
[V_{i,z}-V]^2\rangle}$, with results for the normalized values
$\sigma_v/V$ shown in Fig. \ref{fig:xidv}(b). $\sigma_V/V$ is
independent of $Re$ for $Re\lesssim 0.05$, as expected in the Stokes
regime, and the value $\sigma_V/V\approx 0.75$ is in good agreement
with results found in Stokes flow sedimentation
\cite{nandg2,segreprl,nature}. In a similar way to that seen for
$\xi$ vs. $Re$, the onset of inertial influence occurs at $Re\approx
0.05$, and at higher speeds  the fluctuations sharply decline with
$Re$. At our highest speed,  $Re=2.3$, $\sigma_v/V$ is reduced by
$\approx 80\%$.

We now address whether the dependencies found for the correlation
lengths and the fluctuation magnitudes can be interrelated by a
simple extension of the Hinch model \cite{candl,hinch} that was
originally developed for the  Stokes regime
\cite{hinch,segreprl,nature}. Within this model, velocity
fluctuations arise from particle density fluctuations in regions
whose spatial extent is  the correlation length $\xi$.
 For a {\it random} particle
 configuration, the average number of particles in a
region of size $\xi$ is $N_\xi=\xi^3\phi/v_p$, where $v_p=4\pi
a^3/3$. The fluctuations in number are
 $\Delta N_\xi=\sqrt{N_\xi}$, and mass
$\Delta m_\xi=\Delta N_\xi v_p(\rho_{particle}-\rho_{fluid})$. In
steady state, the buoyancy force acting on these regions,
$F_g=\Delta m_\xi g$, is equal to the viscous drag force
$F_D=6\pi(1+\beta_{Re})\eta\xi\Delta V$,
 yielding $\Delta V=\Delta m_\xi g/6\pi(1+\beta_{Re})\eta\xi$.
 The term $\beta_{Re}\approx
0.133Re^{0.78}$ is the inertial part of the drag force,
 valid for $Re\lesssim 30$ \cite{leclair}.
Additionally, because the drag force applies to  regions of size
$\xi$, the appropriate Reynolds number is
$Re_\xi=2V\xi/\nu=(\xi/a)Re$. The modified Hinch model then becomes
\begin{equation}
\sigma_V/V\approx 0.6 [1+0.133([\xi/a]
Re)^{0.78}]^{-1}\sqrt{\xi\phi/a}, \label{eq:dvre}
\end{equation}
where the bracketed term represents the influence of particle
inertia, and the Hinch model is recovered when $Re\rightarrow 0$
\cite{nature}. (We also neglect a prefactor \cite{nature},
$\gamma(\phi)\equiv
\frac{V_0}{V(\phi)}\frac{\eta_0}{\eta(\phi)}\sqrt{S(\phi,0)}\approx
1.0$ in our semi-dilute samples).

 To test this model, we input our
fit expression for $\xi$ from Eq. (\ref{eq:xifit}) into Eq.
(\ref{eq:dvre}) and directly compare the predicted values for
$\sigma_v/V$ with our data. As seen  in Fig. \ref{fig:xidv}(b), the
agreement over the entire range of Reynolds numbers is remarkably
good, particularly considering the relative simplicity of the model.
The agreement shows that the drop in fluctuation magnitudes with
$Re$ is due to both (i) the decreasing correlation lengths $\xi$ and
(ii) the increasing drag force term $\beta_{Re}$. To illustrate
their relative importance, Fig. \ref{fig:xidv}(b) shows the model
 predictions  when the inertial drag force
corrections are neglected ($Re=0$, dashed line). The result shows
only a slight decrease with $Re$, now due solely to the reduction in
$\xi$, and greatly underestimates the observed drop in fluctuation
magnitudes. This shows that both terms (i) and (ii) are of similar
importance in capturing the behavior of  fluctuations at moderate
Reynolds numbers.
\begin{figure}[t]\includegraphics[width=80mm]{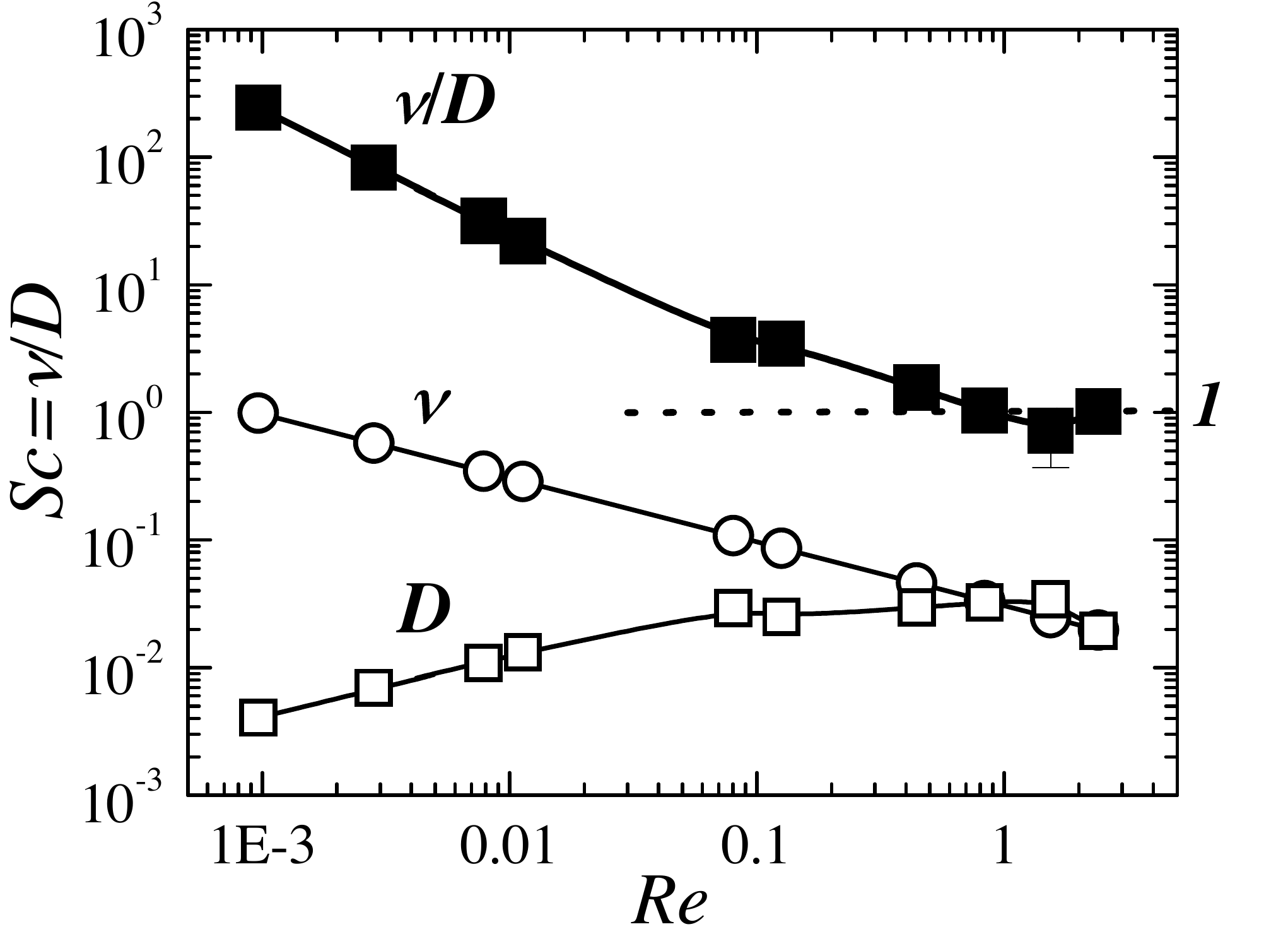}
\caption{Fluid kinematic viscosity $\nu$, particle diffusion
coefficient $D=0.4\sigma_V \xi$, and their ratio $\nu/D$ (the
Schmidt number $Sc$) vs. $Re$. The sedimentation dynamics change
when $D$ becomes as large as $\nu$. The solid lines are a guide to
the eye.} \label{fig:sc}
\end{figure}

The  value of the Reynolds number, $Re\approx 0.05$, at which
inertial effects first appear can be
 rationalized in two different ways. First, it seems reasonable to expect that
 when the inertial screening length $l$
 becomes similar to the Stokes correlation length $\xi_0$,
 the fluctuations will  show an inertial
 influence. This translates to an onset
 value $l=a/Re_c\sim \xi_0$, or $Re_c\sim a/\xi_0\approx 0.04$, in excellent agreement
 with our estimate of $Re\sim 0.05$ from Fig. \ref{fig:xidv}.

Interestingly, Brenner  proposed  a different criterion   based
solely upon a comparison of the particle diffusion coefficient $D$
and the solution viscosity $\nu$ \cite{michael}. Particle diffusion
is driven by velocity fluctuations, and can be estimated from
 $D=0.4 \xi \sigma_V$ \cite{nandg2,nature,nissila}. In the viscous Stokes regime
  $D\ll \nu$. Brenner argued that if the viscosity is reduced
to the point where particle diffusion becomes similar to (viscous)
momentum diffusion, $D\sim \nu$, particles will diffuse faster than
the momentum they are releasing into the fluid, so that the purely
viscous Stokes flow conditions no longer apply. To test this
criterion, we plot in Fig. \ref{fig:sc} our values for $\nu$, $D$,
and the ratio  $\nu/D$. The results generally confirm the Brenner
picture.  The  region where $\nu/D\sim 1$ is indeed where
significant deviations to Stokes behavior are seen.

Finally, to demonstrate the generality of our results, we re-examine
the experiments and conclusions of Cowan et {\it al.} \cite{john}.
They compared many concentrations $0.19\le \phi\le 0.5$, at both
$Re_0=0.007$ and $Re_0=0.3$, and concluded that, contrary to
expectations, the velocity fluctuations were independent of Reynolds
number for $Re<1$. We first note that they defined $Re_0\equiv
2aV_0/\eta_0$ based upon the infinite dilution values $V_0$ and
$\eta_0$, not the values at the high concentrations studied. Using
our definition $Re=2aV/\eta$, we estimate that the $Re_0=0.3$
samples range from $0.001\lesssim Re \lesssim 0.05$. Our critical
Reynolds number for the appearance of inertial influence, $Re_c=
a/\xi_0$, can also be evaluated from their findings that $\xi\sim
11a\phi^{-1/3}$, resulting in $Re_c\approx 0.06$. We therefore find
that $Re\lesssim Re_c$ for {\it all} of their samples, consistent
with their findings of no inertial effects..

    The results described here show that the onset of inertial
    corrections to Stokes sedimentation occurs when the  inertial screening length
    $l=a/Re$ becomes as small as the Stokes
    velocity correlation length $\xi_0$, at  Reynolds number
    $Re_c=a/\xi_0$. At higher Reynolds numbers, inertial screening
    reduces the size and magnitude of the velocity correlations.
    The success of the Hinch model in connecting $\sigma_V/V$ with
    $\xi$  suggests that the models  underlying
    assumption of a random particle density distribution remains
    valid in the presence of inertia.
    These results provide an important benchmark for future
    theoretical work on moderately inertial systems.

We thank Tony Ladd for many fruitful discussions.

\end{document}